\documentclass[aps,prl,twocolumn,superscriptaddress]{revtex4-1}

\usepackage{amsmath}
\usepackage{amsfonts,amssymb, graphicx}
\usepackage{wrapfig}
\usepackage{afterpage}
\usepackage{natbib}
\usepackage{color}
\usepackage[usenames,dvipsnames,svgnames,table]{xcolor}
\usepackage{braket}
\usepackage{stmaryrd}
\usepackage{tikz}
\usepackage{pgffor}
\usepackage{verbatim}
\usepackage{float}
\usepackage{mathrsfs}
\usepackage{soul}
\usepackage[normalem]{ulem}
\usepackage[colorlinks,breaklinks=true,linkcolor=blue, citecolor=blue,urlcolor=blue, linktocpage=true]{hyperref}

\newcommand{\be}{\begin{equation} }
\newcommand{\ee}{\end{equation} }
\newcommand{\ba}{\begin{eqnarray} }
\newcommand{\ea}{\end{eqnarray} }

\newcommand{\bpm}{\begin{pmatrix}}
	\newcommand{\epm}{\end{pmatrix}}
\newcommand{\bmm}{\begin{matrix}}
	\newcommand{\emm}{\end{matrix}}

\newcommand{\bea}{\begin{eqnarray}}
\newcommand{\eea}{\end{eqnarray}}

\newcommand{\heff}{\hbar_{\rm eff}}
\newcommand{\tilambda}{\tilde{\lambda}}
\newcommand{\Ccl}{C_{\rm cl}}

\newcommand{\llangle}{\left<\hspace{-5.83pt}\left<}
\newcommand{\rrangle}{\right>\hspace{-5.83pt}\right>}

\begin{document}
	\title{ Early-Time Exponential Instabilities in Non-Chaotic Quantum Systems}
	
	\author{Efim B. Rozenbaum}
	\email[]{efimroz@umd.edu}
	\affiliation{Joint Quantum Institute, University of Maryland, College Park, MD 20742, USA.}
	\affiliation{Condensed Matter Theory Center, Department of Physics, University of Maryland, College Park, MD 20742, USA} 
	\author{Leonid A. Bunimovich}
	\affiliation{School of Mathematics, Georgia Institute of Technology, Atlanta, GA 30332, USA}
	\author{Victor Galitski}
	\affiliation{Joint Quantum Institute, University of Maryland, College Park, MD 20742, USA.}
	\affiliation{Condensed Matter Theory Center, Department of Physics, University of Maryland, College Park, MD 20742, USA} 
	\begin{abstract}
    	The vast majority of dynamical systems in classical physics are chaotic and exhibit the butterfly effect: a minute change in initial conditions can soon have exponentially large effects elsewhere. But this phenomenon is difficult to reconcile with quantum mechanics. One of the main goals in the field of quantum chaos is to establish a correspondence between the dynamics of classical chaotic systems and their quantum counterparts. In isolated systems in the absence of decoherence, there is such a correspondence in dynamics, but it usually persists only over a short time window, after which quantum interference washes out classical chaos. We demonstrate that quantum mechanics can also play the opposite role and generate exponential instabilities in classically non-chaotic systems within this early-time window. Our calculations employ the out-of-time-ordered correlator (OTOC) -- a diagnostic that  reduces to the Lyapunov exponent in the classical limit, but is  well defined for general quantum systems. Specifically, we show that a variety of classically non-chaotic models, such as polygonal billiards, whose classical Lyapunov exponents are always zero, demonstrate a Lyapunov-like exponential growth of the OTOC at early times with Planck's-constant-dependent  rates. This  behavior is sharply contrasted with the slow early-time growth of the analog of the OTOC in the systems' classical counterparts. These results suggest that classical-to-quantum correspondence in dynamics is violated in the OTOC even before quantum interference develops.
	\end{abstract}
	
	\maketitle
	
	{\it Introduction ---} Quantum mechanics has a general effect that it washes out sharp features of classical dynamics due to its wave-like nature and the uncertainty principle. This effect becomes crucial for chaotic systems because sharp features such as sensitive dependence on initial conditions, that is the butterfly effect, are eventually destroyed. In isolated systems, this suppression of the butterfly effect occurs after a  short period of semiclassical evolution -- the length of this period grows logarithmically with system size~\cite{Berman78, *Zaslavsky81, Toda87, *Gu90, *zurek1994decoherence, *Zurek95, Zurek98, Berry01}. This time scale is known as the scrambling or Ehrenfest time, $t_E$.
	
Even though the scrambling  time is usually  short even  in macroscopic isolated systems  (which seems to be in a disagreement with observable phenomena), system's decoherence often resets dynamics back to the semiclassical regime. This explains why classical chaotic dynamics is ubiquitously observable~\cite{Zurek98, Zurek99, Berry01, Schlosshauer08} (for alternative  views on the long Erhenfest-time ``paradox,'' see  Refs.~\cite{Wiebe05, *Ballentine08}). Regardless of the explanation, the behavior of quantum systems in the Erhenfest window and the fate of classical-to-quantum correspondence in this regime are clearly of fundamental interest, and we focus on this regime in the present manuscript.
	
	\begin{figure} 
		\includegraphics[width=\linewidth]{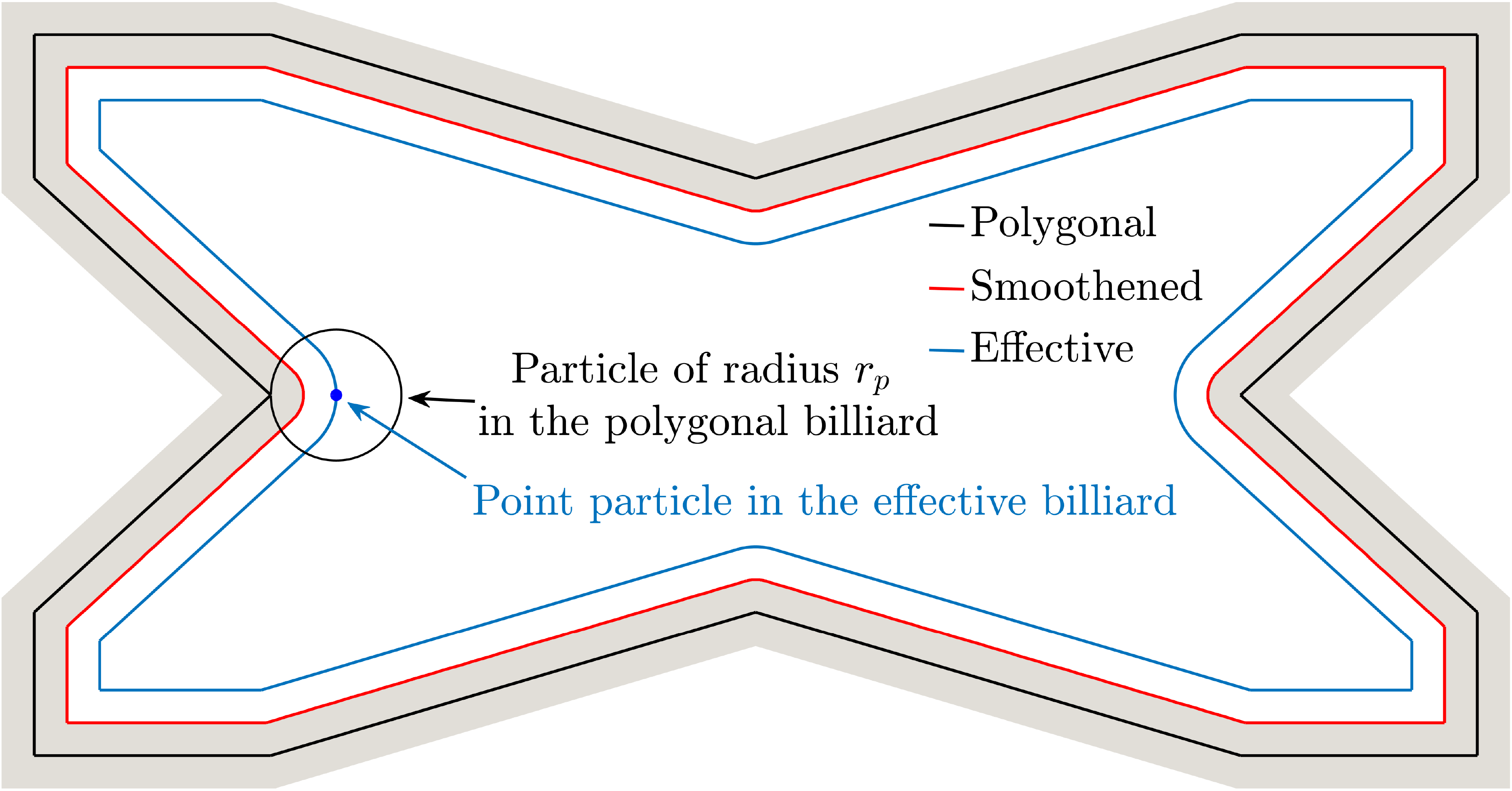}
		\caption{Outer black line: polygonal butterfly-shaped billiard. The area is unit. Inner blue line: effective mathematical billiard hosting a point particle classically equivalent to the outer polygonal billiard hosting a rigid circular particle of radius $r_p = \sigma\sqrt{\heff/2}$ and zero moment of inertia. Note that the inward-pointing corners of the polygonal billiard are rounded into circular arcs or radius $r_p$, making the effective mathematical billiard classically chaotic with positive Lyapunov exponent. Gray shaded region: a close sub-$r_p$ vicinity of the billiard wall: small changes of the billiard geometry within this region do not affect the early-time quantum dynamics. Middle red line: a smoothened billiard used for comparison purposes below.}\vspace{-5pt} \label{fig:ButtBill}
	\end{figure}
	
	In particular, we demonstrate here that in contrast to conventional wisdom, quantum mechanics can induce certain short-time exponential instabilities in models, which are classically non-chaotic. While our construction, described below, is specific to billiards and out-of-time-ordered correlators, we believe that this behavior can exist in a variety of dynamical systems. Besides, both classical and quantum billiards are deeply connected to disordered metals~\cite{Gorkov65, *Efetov83, *Altshuler86, *Doron92, *Serota92, *Argaman93} (for a review, see Ref.~\cite{Montambaux95}), transport phenomena in various systems, such as propagation of particles through rippled channels (see, e.g., Refs.~\cite{Jalabert90, *Baranger91, *Luna-Acosta96, *Herrera-Gonzalez11} and references therein), and also quantum dots (see, e.g., Ref.~\cite{QuantumDotsBilliardsBook}). 
	
	We start with a model based on an illustrative set of observations. Consider a classical ``mathematical billiard,'' i.e. a point particle within a closed domain reflecting off of its hard walls, such as the polygonal black shape in Fig.~\ref{fig:ButtBill}. It has been rigorously proven~\cite{Zemlyakov75, *Boldrighini78} that the Kolmogorov-Sinai (KS) entropy and the closely related Lyapunov exponents of {\em any} polygonal billiard are strictly zero. Next, consider the corresponding ``physical billiard,'' a classical hard disk of radius $r_p$ reflecting off of the same polygonal walls. Clearly, this physical billiard is equivalent to a mathematical billiard of a smaller size, since the particle's center is not allowed to approach the walls of the physical billiard closer than $r_p$. Such equivalent billiard is shown by the inner blue shape in Fig.~\ref{fig:ButtBill}. We assume that the particle's mass is concentrated in the center, and ignore rotational motion. A crucial observation~\cite{Bolding18, *Bunimovich19} is that this redrawing may give rise to a smoothing of sharp features of non-convex polygons, such as the black shape in Fig.~\ref{fig:ButtBill}. The resulting shape is no longer a polygon, and the obstruction for the KS entropy to vanish is removed. Indeed, the inner blue billiard in Fig.~\ref{fig:ButtBill} is classically chaotic, with a positive Lyapunov exponent. Finally, consider a quantum particle embedded into a non-convex polygonal billiard. Semiclassical early-time dynamics of a quantum wave packet is in a certain sense similar to motion of a finite-size classical particle; i.e., classically chaotic motion in the physical billiard. As shown below, there is indeed the onset of exponential instabilities in the classically non-chaotic systems such as this one, hence providing an example of violation of the conventional view on the classical-to-quantum correspondence. 
	
	\begin{figure} 
		\includegraphics[width=\linewidth]{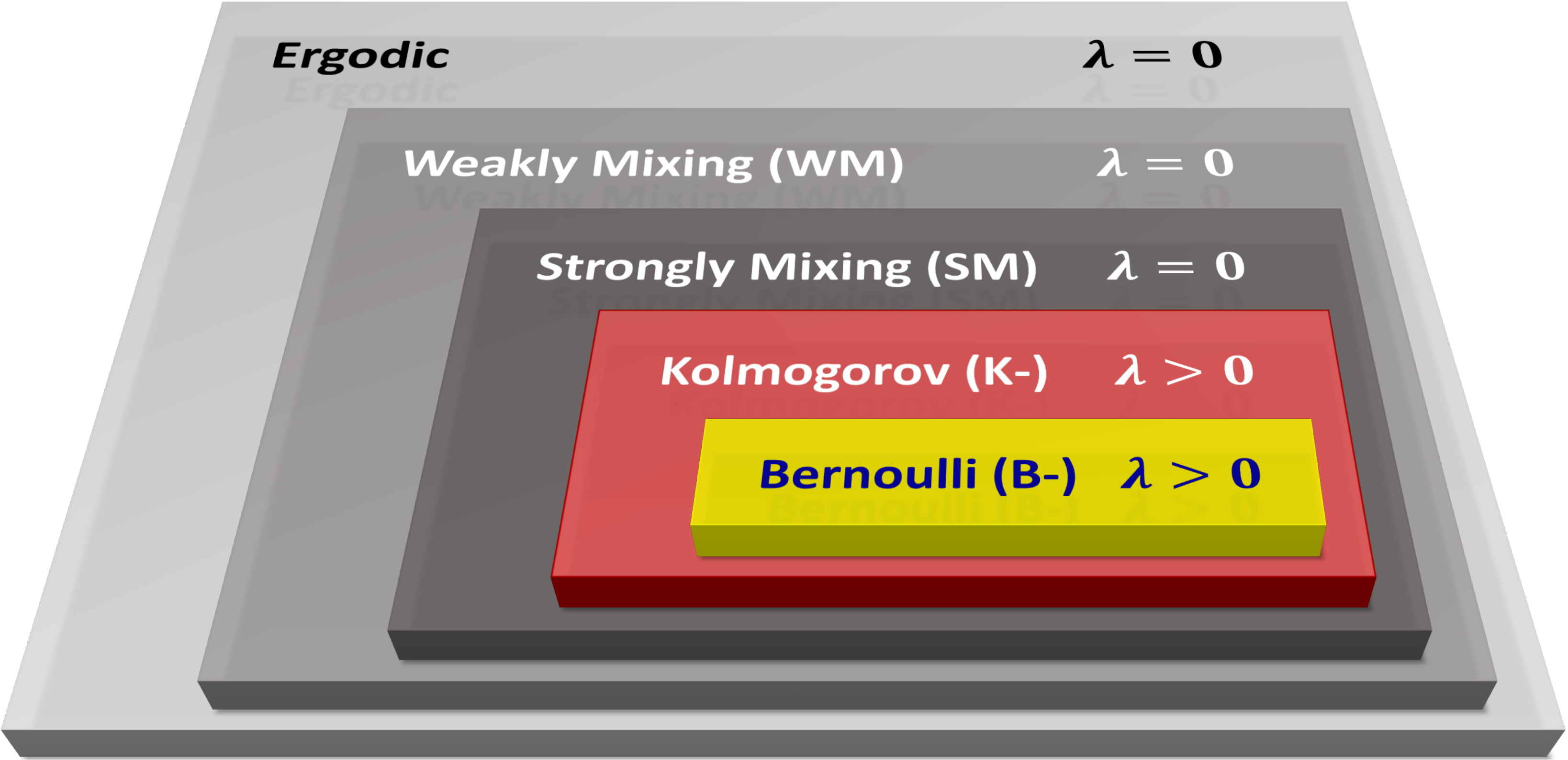}
		\caption{Ergodic hierarchy~\cite{Sinai77ErgTheoryBook} provides a nested classification of non-integrable systems. Only K- and B-systems are chaotic and have positive Lyapunov exponents, while merely ergodic and merely mixing systems have no exponential instabilities.}\vspace{-5pt} \label{fig:ErgHier}
	\end{figure}
	To diagnose this behavior, we employ the out-of-time-ordered correlator (OTOC). The OTOC was introduced by Larkin and Ovchinnikov~\cite{Larkin69} in the context of quasiclassical approximation in the theory of superconductivity in disordered metals and used recently in the pioneering works by Kitaev~\cite{kitaev} and Maldacena et al.~\cite{Maldacena16} to define and describe many-body quantum chaos with an eye on fundamental puzzles in black-hole physics. In the last few years, the OTOC has become a popular tool to describe ``quantum chaos'' in many-body quantum systems (see e.g. Refs.~\cite{Roberts15, *Roberts16, *huang2016out, *fan2016out, *chen2016quantum, *swingle2016measuring, *swingle2016slow, *yao2016interferometric, *Mezei17, *Syzranov17, *Khemani18, *Xu18}). It was shown in Refs.~\cite{Rozenbaum17,ChavezCarlos19,Rozenbaum18} that the exponential growth of the OTOC, although not always equal, might be connected to the exponential divergences of orbits in the phase space of an effective classical system. In certain cases, such as the celebrated Sinai billiard~\cite{Sinai70} and Bunimovich stadium~\cite{Bunimovich74, *Bunimovich79}, it is straightforward to understand this classical limit. Below, we consider non-chaotic polygonal billiards instead. In a polygon, for any pair of trajectories -- no matter how close the initial conditions are -- one can identify the origin of each trajectory evolving the dynamics backward in time~\cite{Zemlyakov75, *Boldrighini78}, ensuring that the KS entropy is zero. Note that in the ergodic hierarchy, which is displayed in Fig.~\ref{fig:ErgHier} in the order of ``increasing chaoticity,'' polygonal billiards fall within at most the strongly mixing class (only K- and B-systems have a positive KS entropy; see e.g. Ref.~\cite{Sinai77ErgTheoryBook} for a detailed discussion of the hierarchy). Interestingly, however, the mixing property at the classical level can be sufficient to generate Wigner-Dyson or intermediate energy-level statistics on the quantum side, as was shown, for example, in Ref.~\cite{Lima13} for a family of irrational triangular billiards~\cite{Casati99}.
	
	Apart from this ``quantum'' Lyapunov instability, where quantum mechanics effectively promotes the corresponding classical system in the ergodic hierarchy, there are potentially more prosaic sources of early-time instabilities in OTOC in various systems. First, note that the classical definition of exponential Lyapunov instabilities involves taking two limits: infinitesimally small initial separation in the phase space and infinite time-limit in the subsequent evolution. However, neither limit is available quantum-mechanically because a wave-packet always has a finite size per uncertainty principle and subsequently spreads out on time-scales of order the Ehrenfest time. Second, there is a  distinction between the quantum-mechanical expectation value in the way quasiclassical trajectories are accounted for and the classical phase-space average (See, e.g., Ref.~[\onlinecite{Rozenbaum17}]). Therefore, in most numerical simulations of OTOCs the proper Lyapunov limit can not be enforced and the dynamics of the wave-packets may involve rapid growth, which is however spurious in nature. To explore these types of phenomena, we also study convex polygonal billiards (specifically an irrational triangle) and some integrable systems.
		
\begin{figure}
		\begin{minipage}[c]{0.5\linewidth}
			\includegraphics[width=\textwidth]{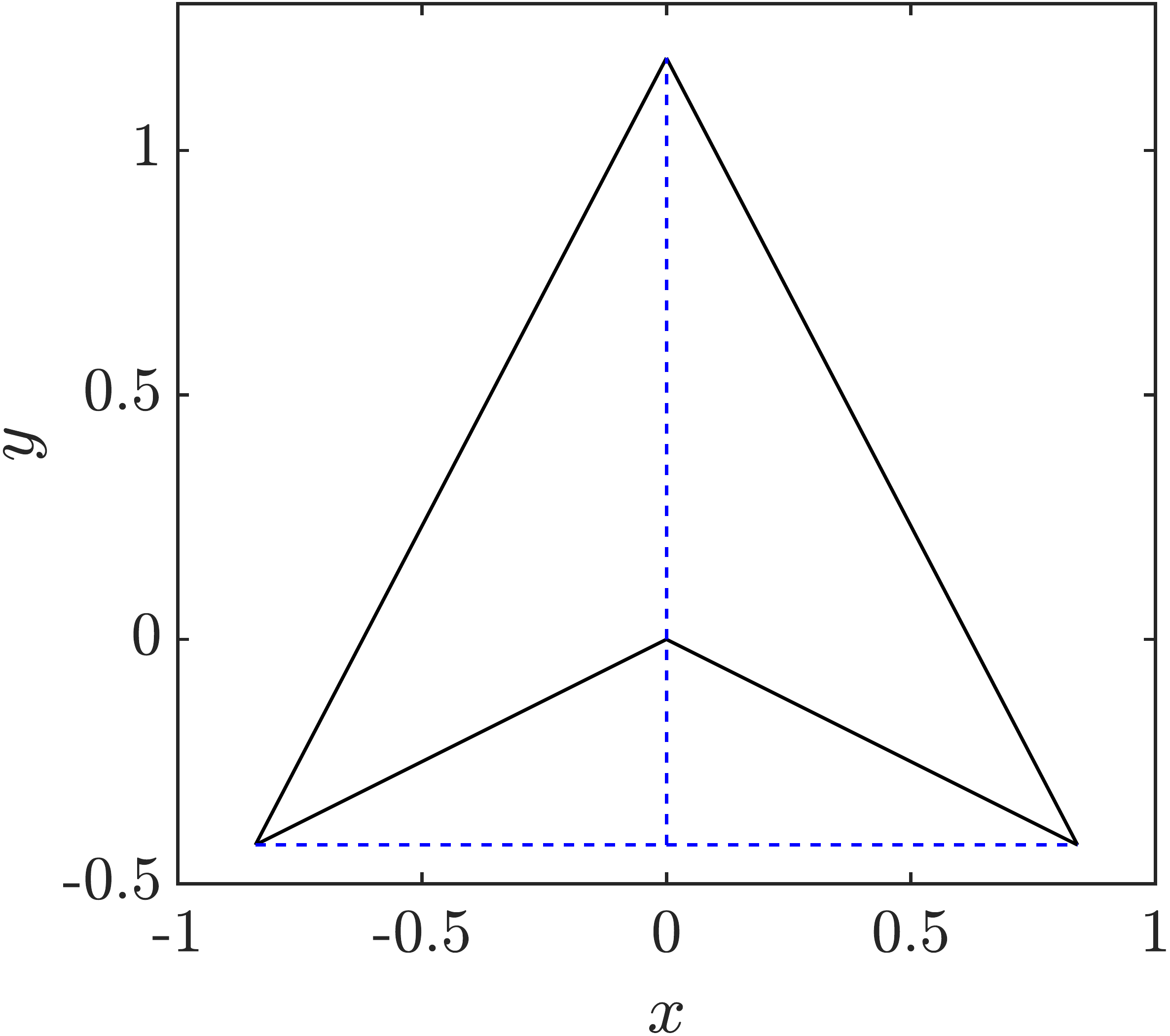}
		\end{minipage}\hfill
		\begin{minipage}[c]{0.445\linewidth} \vspace{-20pt}
			\caption{``Deformed triangular'' (quadrilateral) billiard. For a finite-sized particle, the inward-pointing corner gets rounded in the same way as those in the butterfly-shaped billiard in Fig.~\ref{fig:ButtBill}.} \label{fig:TriBill}
		\end{minipage}
		\vspace{-6pt}
	\end{figure}

	{\it Models ---} We perform explicit calculations for the butterfly-shaped polygonal billiard shown in Fig.~\ref{fig:ButtBill} (outer black line), the quadrilateral non-convex billiard shown in Fig.~\ref{fig:TriBill}, and a triangular billiard obtained from it by removing the vertex at $(0;0)$.
	
	We launch a wave-packet with the initial wave-function 
	\begin{equation} \label{eq:InitialState}
	\Psi_0({\bf r}) \propto \exp{ \left[ -\frac{ ({\bf r} - {\bf r}_0)^2}{2\heff\sigma^2} + \frac{i}{\heff} {\bf p_0}\cdot {\bf r} \right]}
	\end{equation}
	by decomposing it into the billiard's energy eigenstates and evolving accordingly. This requires numerical solution of the Schr{\"o}dinger equation for the billiard:
	\begin{equation}
	\label{SE}
	\hspace{-0.7pt} -\dfrac{\heff^2}{2}\nabla^2 \Psi({\bf r}) = E\Psi({\bf r}),\,\,\  \Psi({\bf r})\Bigl|_{{\bf r}\in {\rm billard\, walls}} = 0.
	\end{equation}
	Here $\heff = \hbar/(p_0\sqrt{A})$, $A$ is the billiard's area, and $p_0 = |{\bf p}_0|$ is the wave-packet's average momentum. $A=1$ and $p_0=1$ are chosen as the units along with the particle's mass $m=1$. The butterfly-shaped billiard has two reflection symmetries with respect to $x \to -x$ and $y \to -y$. Thus, its eigenstates fall into four parity classes. In order to enforce these parities and speed up the calculations, one typically solves the eigenvalue problem on a quarter of the billiard imposing the Dirichlet and/or Neumann boundary conditions on each cut, thereby determining the parity class of the solutions. We solve these four boundary-value problems for the Laplace operator numerically using the finite-element method, and find  eigenstates of each class up to a certain energy cutoff. The accuracy of the numerical solution generally decreases with the number of found eigenstates \cite{Heuveline03}. We use Weyl's formula for the number of modes \cite{Baltes78} to control it. Weyl's law sets the asymptotic behavior of the average number $\mathcal{N}(E)$ of eigenstates below energy $E$ as: $\mathcal{N}(E) \simeq \left[A/(4\pi)\right]\varepsilon - \left[P/(4\pi)\right]\sqrt{\varepsilon}, \quad \varepsilon\to\infty$, where $\varepsilon=2E/\heff^2$ and $P$ is the billiard's perimeter.  For our present purposes, it is sufficient to use around $N_{\rm max}=10^4$ eigenstates, and within this range, we have exact agreement with Weyl's law, i.e. the number of found states is centered around Weyl's asymptote. In addition, we repeat the calculations with the boundary-integral method and obtain the same results.
	
\begin{figure} 
		\includegraphics[width=\linewidth]{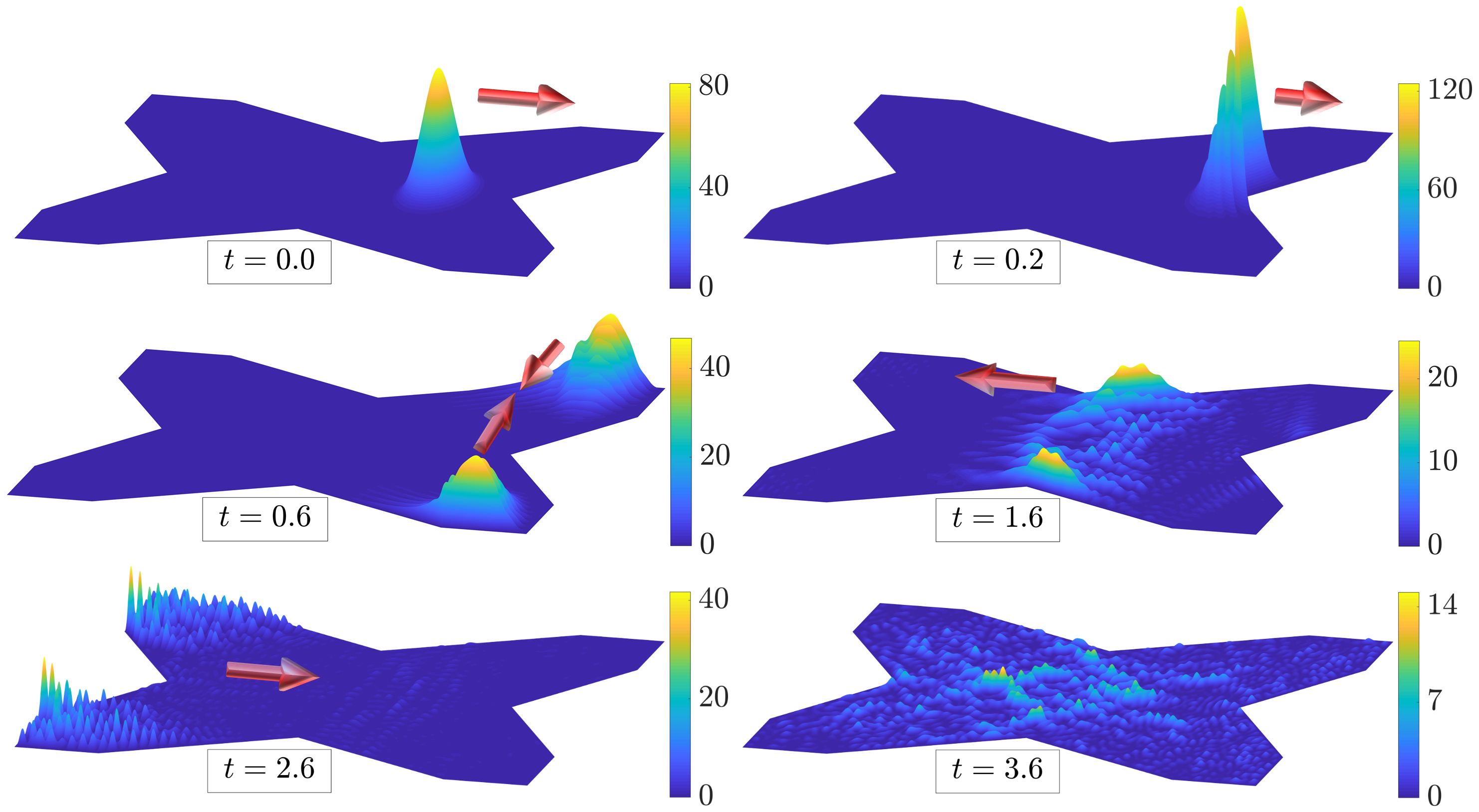}
		\caption{An example of successive stages of the wave-packet evolution, $\left|\Psi({\bf r},t)\right|^2$, in the butterfly-shaped polygonal billiard. Red arrows indicate the directions of motion of the components. Initial velocity is aimed at an inner corner.} \label{fig:Evolution}
		\vspace{-7pt}
	\end{figure}

	Due to the lack of narrow outer corners, the butterfly-shaped billiard allows for a relatively long lifetime of the initial minimal-uncertainty wave packet until this packet becomes completely scrambled and loses classical-like dynamics (see Fig.~\ref{fig:Evolution}). Along with this billiard, we introduce an effective mathematical billiard (Fig.~\ref{fig:ButtBill}, inner blue line) that is obtained by tracing the set of positions available to the center of a circular particle of radius $\sigma\sqrt{\heff/2}$ inside the polygonal butterfly billiard. The squeezing parameter $\sigma$ is defined in Eq.~(\ref{eq:InitialState}). 
	
	{\it Diagnostic tool ---} As a measure of quantum chaotic dynamics, we use the OTOC~\cite{kitaev, Maldacena16, Roberts15, *Roberts16, *huang2016out, *fan2016out, *chen2016quantum, *swingle2016measuring, *swingle2016slow, *yao2016interferometric, *Mezei17, *Syzranov17, *Khemani18, *Xu18, Rozenbaum17, ChavezCarlos19, Rozenbaum18} defined as: \vspace{-3pt}
	\begin{equation}
	\label{L}
	C(t) = - \left<\left[\hat{x}(t),\, \hat{p}_x(0) \right]^2\right>,\vspace{-3pt}
	\end{equation}
	where $\hat{x}(t)$ and $\hat{p}_x(t)$ are the Heisenberg operators of the $x$-components of the particle's position and momentum. As was first pointed out by Larkin and Ovchinnikov~\cite{Larkin69}, the OTOC probes the sensitivity of quasiclassical trajectories to initial conditions as \mbox{$\hat{p}_x(0)=- i\heff \partial/\partial x(0)$}, and hence \mbox{$C(t)=\hbar^2 \left\langle \left( \frac{\partial x(t)}{\partial x(0)} \right)^2 \right\rangle$}. Therefore, classical Lyapunov-like growth is anticipated at early times, \mbox{$C(t) \propto \exp(2 \tilambda t)$}, for a chaotic system, with $\tilambda$ related to its Lyapunov exponent in the respective subspace. 

	\begin{figure} 
	\includegraphics[width=\linewidth]{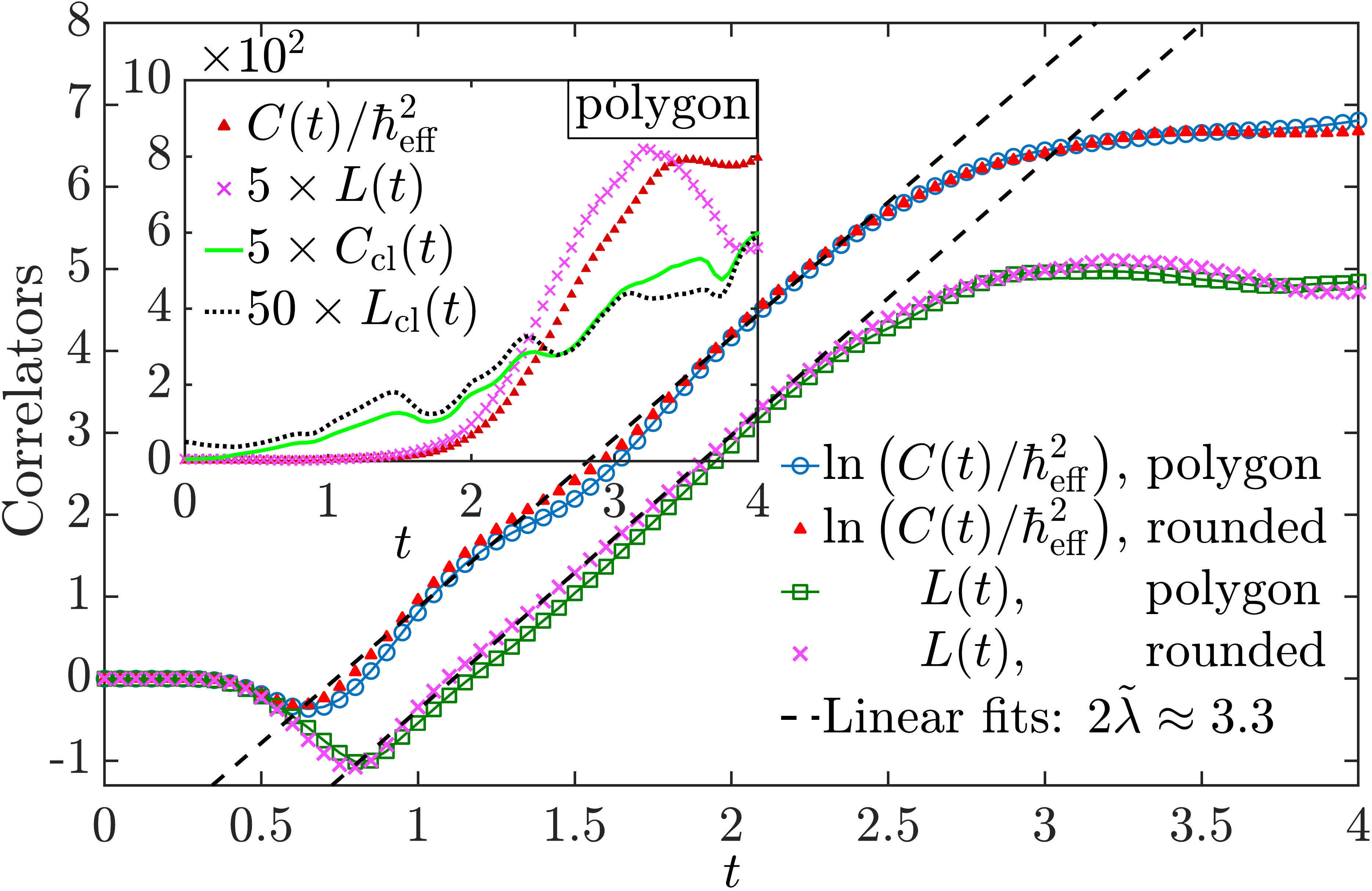}
	\caption{Main plot -- open blue circles and line: logarithm of the OTOC in the polygonal butterfly-shaped billiard: $\ln\left(C(t)/\hbar_{\rm eff}^2\right) = \ln\left(-\frac{1}{\hbar_{\rm eff}^2}\left<\left[\hat{x}(t), \hat{p}_x(0)\right]^2\right>\right)$. Solid red triangles: the same in the rounded version of this billiard (middle red line in Fig.~\ref{fig:ButtBill}). A remarkable agreement demonstrates that the growth in both cases is the same, supporting our finite-size-related arguments. In addition, we show the corresponding behavior of an alternative diagnostic, $L(t) = \left<\ln\left(-\frac{1}{\hbar_{\rm eff}^2}\left[\hat{x}(t), \hat{p}_x(0)\right]^2\right)\right>$, that swaps the order of averaging and logarithm to that of the proper definition of the classical Lyapunov exponent. For chaotic systems with uniform phase space, one would expect \mbox{$L(t) = 2\tilambda t + \rm{const}$} at $t<t_E$. Green squares and line: $L(t)$ in the polygonal butterfly-shaped billiard. Pink crosses: $L(t)$ in the rounded billiard. Dashed black lines: linear fits for $\ln(C(t)/\heff^2)$ and $L(t)$ in the polygon. Both show the exponent $2\tilambda \approx 3.3$ that is 5 times larger than the inverse time-window, which ensures that the fit is valid. Inset -- the comparison between $C(t)/\heff^2$ and $\Ccl(t) = \llangle\,\left\{x(t), p_x(0)\right\}_{\rm Poisson}^2\,\rrangle$ [see Eq.~(\ref{eq:Ccl})] and between $\exp\left[L(t)\right]$ and $\exp\left[L_{\rm cl}(t)\right] = \exp\left[\,\llangle\,\ln\left\{x(t), p_x(0)\right\}_{\rm Poisson}^2\,\rrangle\,\right]$ in the polygonal quantum and classical billiards, respectively. $\heff = 2^{-7}$, $\sigma = 1/\sqrt{2}$, $R_s = \frac{\sqrt{2}-1}{16\sqrt{2}}\approx 0.02$.}\vspace{-10pt} \label{fig:OTOCexp_poly_vs_smooth}
\end{figure}

	As was shown in Ref.~\cite{Rozenbaum18}, whether the OTOC actually  grows exponentially or not, depends on an initial quantum state and on the existence of a finite time window between the first collision and the Ehrenfest time. For billiards, a natural choice of the initial state is the minimal-uncertainty wave-packet, Eq.~(\ref{eq:InitialState}). The scrambling (Ehrenfest) time in chaotic systems is short and grows logarithmically slowly with system size: $t_E = \ln(\heff^{-1})/\lambda_{\rm cl}$, where $\lambda_{\rm cl}$ is the positive Lyapunov exponent of the classical counterpart of the system~\cite{Berman78, *Zaslavsky81}. This estimate is based on the fact that, in contrast to non-chaotic systems where the spreading of wave-packets is algebraic in time, the spreading is typically exponential in chaotic systems, i.e. in quantum counterparts of K- and B-systems from the ergodic hierarchy. Extending the Ehrenfest window to cover the long-time ergodic classical behavior, which is required to define the global Lyapunov exponents in chaotic systems, is an exponentially demanding numerical task. However, local finite-time Lyapunov exponents can be defined, although they fluctuate at these short times~\cite{Rozenbaum18}.

	{\it Breakdown of classical-to-quantum correspondence ---} As shown in Ref.~\cite{Rozenbaum18}, in  quantum billiards, which are classically chaotic, the exponential growth of the OTOC may be related to the classical Lyapunov instability  and extends up until the Ehrenfest time. After that, the wave packet is spread across the entire system, and no further exponential growth is possible. 
	
	The classical counterpart of the OTOC is defined as:\vspace{-2pt}
	\begin{equation} \label{eq:Ccl}
	\Ccl(t) = \llangle\lim\limits_{\Delta x(0) \to 0} \left(\frac{\Delta x(t)}{\Delta x(0)}\right)^2\rrangle, \vspace{-2pt}
	\end{equation}
	where $\;\llangle\;\ldots\;\rrangle\;$ denotes the classical phase-space average over the Gaussian Wigner function corresponding to the initial quantum packet in Eq.~(\ref{eq:InitialState}), and $\Delta x$ is the distance along the $x-$axis between a pair of trajectories starting near some point in the phase space. $\Ccl(t)$ agrees with $C(t)/\heff^2$ all the way up to $t_E$. After that, they deviate from each other. The quantum-mechanical OTOC slows down and eventually saturates, while the classical one continues to grow exponentially. 

\begin{figure} 
	\includegraphics[width=\linewidth]{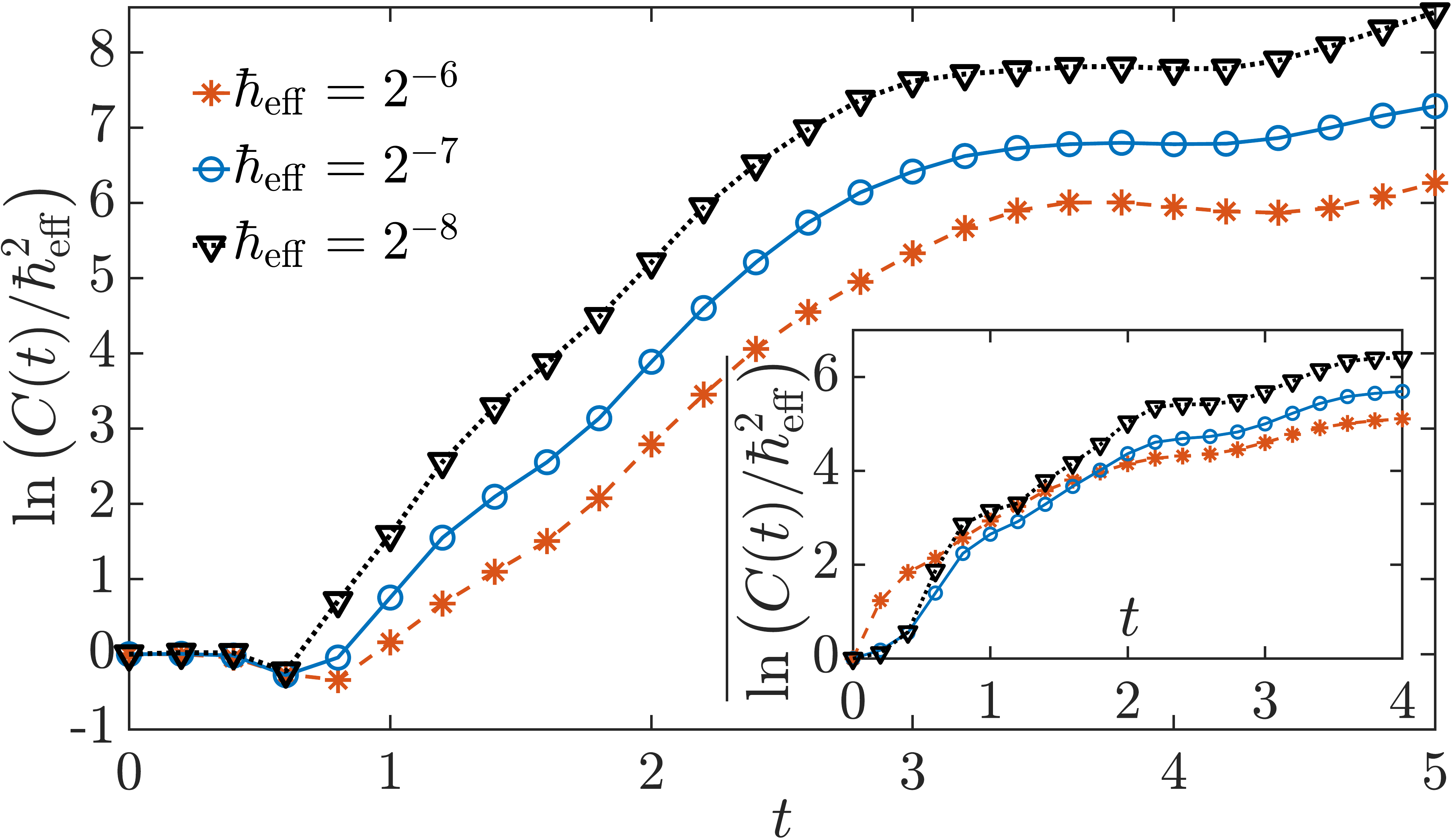}
	\caption{Main plot -- logarithm of the OTOC as a function of time in the polygonal butterfly-shaped billiard at three different values of $\heff$. The exponential growth of the OTOC hinges on the finite wave-packet size. Inset -- logarithm of the OTOC in the quadrilateral billiard (Fig.~\ref{fig:TriBill}), averaged over an ensemble of initial conditions as indicated by the bar, $\overline{\cdot\cdot\cdot}$, with the corresponding values of $\heff$. $\sigma = 1/\sqrt{2}$.}\vspace{-10pt} \label{fig:OTOCexp}
\end{figure}	

\begin{figure} 
		\includegraphics[width=\linewidth]{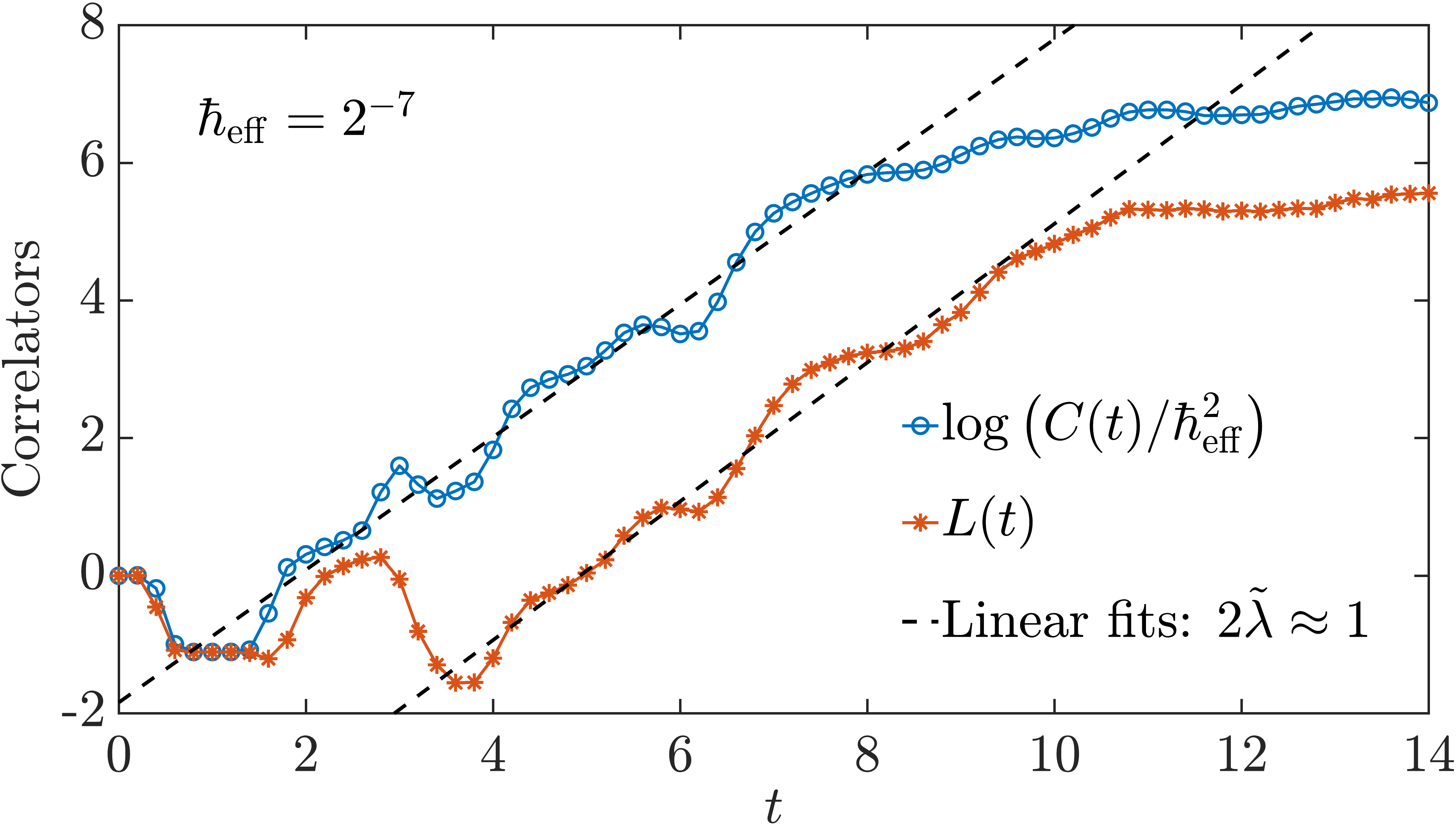}
		\caption{Logarithm of the OTOC, $\ln(C(t)/\heff^2)$, in an irrational triangular billiard (upper blue line and open circles). After an initial-condition-dependent delay, the OTOC shows exponential grows, although at a rate lower than that for the non-convex billiard. The other related correlator, $L(t)$, introduced in Fig.~\ref{fig:OTOCexp_poly_vs_smooth}, is shown for comparison (lower red line and asterisks). Black dashed lines show linear fits with $2\tilambda\approx 1$, which is over 5 times larger than the inverse time window, ensuring that the fit is valid.}\vspace{-10pt} \label{fig:OTOC_regtri}
	\end{figure}

	In the polygonal billiards, there are no positive classical Lyapunov exponents, and the corresponding classical OTOC does not grow exponentially at any time, as shown in the inset in Fig.~\ref{fig:OTOCexp_poly_vs_smooth} for the case of the butterfly-shaped polygonal billiard (dotted black and solid green lines). However the quantum-mechanical OTOC in polygonal billiards shows a clear exponential growth at early times that has no origin in the classical counterparts, as demonstrated in Fig.~\ref{fig:OTOCexp_poly_vs_smooth} (main plot and inset), as well as in Figs.~\ref{fig:OTOCexp}, and \ref{fig:OTOC_regtri} (described below).
	
	As discussed in the introduction, the motion of a minimal-uncertainty wave packet is in some sense similar to that of a finite-size disk. Classical motion of such a disk gives rise to an effective billiard which hosts a point-like particle at the disk's center that is not allowed to approach the walls of the original billiard closer than by the disk's radius. Many billiards preserve their status within the ergodic hierarchy upon this procedure (e.g., a Bunimovich stadium remains a stadium with a smaller area and convex polygons also turn into similar convex polygons). Not so for non-convex polygonal billiards, which go up the ergodic hierarchy for a finite-sized particle from the strongly mixing class to the K-chaotic one. In such non-chaotic systems, there can still be measure-zero sets of unstable points, and these get smeared over finite-measure regions by introducing a finite size of the particle. A quantum wave packet, which always has a finite width, can have a similar effect. 
	
	An additional consideration is that polygons constitute an everywhere dense measure-zero set in the space of closed curves on a plane, and the phase space of the corresponding billiards contains isolated unstable points. A slight variation of the wall's shape almost always results in finite-curvature regions and smears out singular phase-space points. A possible consequence of that would manifest in that a quantum-mechanical wave packet effectively ``rounds'' singularities even if they originate from outer corners of polygons, including those in convex polygonal billiards considered below. This can be generalized to a statement that quantum mechanics promotes measure-zero sets of unstable points into finite-measure sets. We check these conjectures by varying the billiard's boundary within the shaded gray region in Fig.~\ref{fig:ButtBill}, and, in particular, compare the behavior of the OTOC in the polygonal and in a rounded billiard, such as the middle red line in Fig~\ref{fig:ButtBill}. The latter system is classically chaotic. We find a good agreement between the quantum OTOCs in the two, as demonstrated in Fig.~\ref{fig:OTOCexp_poly_vs_smooth}. In addition, from this comparison we can infer that there are no significant effects related to the non-smoothness of the polygonal boundary, such as diffraction, as in the case in the quantum baker's map~\cite{Lakshminarayan19}. Note a major difference between the Lyapunov behaviors of the quantum baker's map and our billiards: the latter do not have a classical Lyapunov exponent at all and the exponential growth of the OTOC there is a purely quantum effect.
	
At smaller values of $\heff$, the wave packets are tighter, and their sides become steeper. Following the reasoning in Refs.~\cite{Prosen02, *Prosen03}, it causes the rate of the OTOC's divergence, $\tilambda$, to be larger than that at larger values of $\heff$, as shown in Fig.~\ref{fig:OTOCexp}, main plot.  The inset in Fig.~\ref{fig:OTOCexp} shows an analogous behavior for the quadrilateral billiard shown in Fig.~\ref{fig:TriBill}.
	
{\it Quantum dynamics in convex polygonal and integrable billiards ---} Classical convex polygonal billiards do not change their status within the ergodic hierarchy upon promoting their point-particle versions to those with finite-size hard particles. However, quantum-mechanically, they still show a rapid initial growth of the OTOC coexisting with an oscillatory behavior, as we demonstrate for an irrational triangular billiard obtained from the quadrilateral one in Fig.~\ref{fig:TriBill} by removing the vertex at $(0;0)$. The effective rate of growth is smaller than for the non-convex billiard, but similar signs of instability are  present. Note that this growth should not be attributed to any mixing dynamics in the classical counterparts of these billiards. Upon quantization, the level statistics of irrational triangular billiards -- the most widely used ``quantum-chaotic'' diagnostic -- is close to the Wigner-Dyson surmise~\cite{Lima13}, putting them outside of the Bohigas-Giannoni-Schmit conjecture~\cite{Casati80, *Bohigas84}. As shown in Fig.~\ref{fig:OTOC_regtri}, the early-time behavior has a period of what looks very much like an exponential growth, although it is modulated by the effects of collisions with the walls.
	
We believe that this behavior of OTOC in convex billiards is due to the fact that a quantum simulation can not as a matter of principle access the proper small-distance and long-time limit where classical Lyapunov exponents are defined. As such, this type of growth in  OTOC is a property of the initial wave-packet rather than the system where it propagates. If so, similar growth should be observable in integrable systems as well. We have considered the simplest billiard --  the rectangular billiard -- and indeed found that a weak growth can be detected (again superimposed on oscillations). Since the rectangle factorizes into two one-dimensional segments, one can also look at the most basic textbook quantum mechanical problem -- a particle in a box. The OTOC can be calculated to a large degree analytically in this case and shows clear recurrent oscillations including short time-intervals of growth. Of course these  periods of growth have no relation to chaos or the butterfly effect and do not contain any valuable information. The behavior of OTOC in these integrable systems is presented in the Supplemental Material~\cite{OnlineSupplement} and will be discussed in more details elsewhere.
	
All in all, there appear to exist two sources of rapid growth of OTOC in billiards: one related to a genuine Lyapunov instability (in chaotic billiards and those that are promoted in the chaotic hierarchy upon quantization) and a spurious growth related to a finite-size wave-packet enforcing the ``wrong'' averaging of the underlying classical dynamics. Such growth is present independently of the status of the effective billiard in the chaotic hierarchy. In order for OTOC to have a physical meaning, it is important to disentangle the two types of contributions. It should be possible by looking at how the growth rates scale with the Planck constant and extracting the ``interesting'' genuine Lyapunov growth, if any. The details of this scaling procedure will be discussed in a subsequent publication. 
	
	\begin{acknowledgments}
		This research was supported by DOE-BES DESC0001911 (V.G.), NSF grant DMS-160058 (L.B.), US-ARO contract No. W911NF1310172 (E.R.), and Simons Foundation (E.R. and V.G.). The authors acknowledge helpful discussions with Sriram Ganeshan, Arul Lakshminarayan, and Meenu Kumari and thank Colin Rylands for proofreading the manuscript. The authors are grateful to Centro Internacional de Ciencias (Cuernavaca, Morelos, Mexico), where this work has been conceived, and personally to Thomas Seligman and Felix Izrailev for outstanding hospitality. \vspace{-10pt}
	\end{acknowledgments}
	
%
	
\end{document}